\documentclass{aastex63}

\usepackage{graphicx}
\usepackage{amsmath}
\usepackage{longtable}
\usepackage{color}
\usepackage{enumerate}

\received{XXX}
\revised{YYY}
\accepted{ZZZ}

\shorttitle{ULXBs and FRBs: Possible Association?}
\shortauthors{Chen et al.}

\begin{document}

\title{Repeating Ultraluminous X-ray Bursts and Repeating Fast Radio Bursts:
A Possible Association?}

\correspondingauthor{Wei-Min Gu}
\email{guwm@xmu.edu.cn}

\author{Hao-Yan Chen}
\affiliation{Department of Astronomy, Xiamen University, Xiamen,
Fujian 361005, P. R. China}

\author{Wei-Min Gu}
\affiliation{Department of Astronomy, Xiamen University, Xiamen,
Fujian 361005, P. R. China}

\author{Jin-Bo Fu}
\affiliation{Department of Astronomy, Xiamen University, Xiamen,
Fujian 361005, P. R. China}

\author{Shan-Shan Weng}
\affiliation{Department of Physics and Institute of Theoretical Physics, 
Nanjing Normal University, Nanjing 210023, P. R. China}

\author{Junfeng Wang}
\affiliation{Department of Astronomy, Xiamen University, Xiamen,
Fujian 361005, P. R. China}

\author{Mouyuan Sun}
\affiliation{Department of Astronomy, Xiamen University, Xiamen,
Fujian 361005, P. R. China}

\begin{abstract}

Ultraluminous X-ray bursts (hereafter ULXBs) are ultraluminous X-ray flares
with a fast rise ($\sim$~one minute) and a slow decay ($\sim$~an hour),
which are commonly observed in extragalactic globular clusters.
Most ULXBs are observational one-off bursts, whereas five flares from
the same source in NGC 5128 were discovered by \citet{Irwin et al.(2016)}.
In this Letter, we propose a neutron star (NS)-white dwarf (WD) binary model
with super-Eddington accretion rates to explain the repeating behavior
of the ULXB source in NGC 5128. With an eccentric orbit, the mass transfer
occurs at the periastron where the WD fills its Roche lobe.
The ultraluminous X-ray flares can be produced by the accretion column 
around the NS magnetic poles.
On the other hand, some repeating fast radio bursts (hereafter FRBs) were also
found in extragalactic globular clusters.
Repeating ULXBs and repeating FRBs are the most violent bursts in 
the X-ray and radio bands, respectively.
We propose a possible association between the repeating ULXBs and
the repeating FRBs.
Such an association is worth further investigation by follow-up observations
on nearby extragalactic globular clusters.

\end{abstract}


\section{Introduction} \label{sec:intro}

Some ultraluminous X-ray bursts (hereafter ULXBs) have been discovered in
recent works \citep[e.g.,][]{Sivakoff et al.(2005),
Jonker et al.(2013),Irwin et al.(2016)}. 
These bursts rise rapidly within one minute to the peak luminosities, 
maintain in a roughly steady ultraluminous state for several hundred seconds,
and then decay to the pre-flare level within a few thousand seconds. 
The peak luminosities of ULXBs ($\ga10^{39} \ \rm{erg \ s^{-1}}$) should be
super-Eddington for a neutron star (NS), which are higher than the peak
luminosities of type-I X-ray bursts
\citep[$\sim 10^{37-38} \ \rm{erg \ s^{-1}}$; e.g.,][]{Galloway et al.(2008)}.
For the Eddington limit of a black hole (BH), it can be estimated from most 
of ULXBs that the peak luminosities are super-Eddington for a stellar-mass BH 
($M_{\rm{BH}} \sim 10 \ M_{\sun}$) or sub-Eddington for an intermediate-mass BH
(IMBH; $M_{\rm{BH}} \sim 10^{2-4}\ M_{\sun}$). Some studies have proposed explanations for ULXBs, such as the shock breakout from a core-collapse
supernova \citep[e.g.,][]{Soderberg et al.(2008)}, the tidal stripping of a
white dwarf (WD) by an IMBH \citep{Shen(2019)}, and the merger of binary NSs \citep[e.g.,][]{Xue et al.(2019)}.

Most of ULXBs are observational one-off bursts, e.g.,  the ULXBs observed in
M86 \citep{Jonker et al.(2013)} and in NGC~4636 
\citep[Source 1 in][]{Irwin et al.(2016)}, whereas two ULXB sources show
repeating behaviors, i.e., two fast flares discovered in NGC~4697
\citep{Sivakoff et al.(2005)} and five flares discovered in NGC~5128 
\citep[Source 2 in][]{Irwin et al.(2016)}. \citet{Sivakoff et al.(2005)}
showed that the ULXBs in NGC~4697 have a peak luminosity of 
$\sim 6 \times 10^{39}\ \rm{erg \ s^{-1}}$, a duration of 
$\sim 70\ \rm{s}$, and a count rate ratio of the flare to the persistent
emission of $\sim 90$. For Source 2 in \citet{Irwin et al.(2016)}, four flares
were revealed by the $Chandra$ data, and the fifth flare was observed by
$XMM-Newton$. The luminosities of persistent emission and peak flare are
$\sim 4 \times 10^{37} \ \rm{erg \ s^{-1}}$ and $\sim 8 \times 10^{39} \ 
\rm{erg \ s^{-1}}$, respectively, which implies an increase of a factor of 
$\sim 200$ within a minute. The optical counterpart is either a massive
 globular cluster \citep[called GC~0320; e.g.,][]{Harris et al.(1992)}, or an
ultra-compact dwarf companion galaxy of NGC~5128 \citep{Irwin et al.(2016)}.

\citet{Maccarone(2005)} suggested that the ULXBs discovered in NGC~4697
could be produced by the accreting eccentric binaries in globular clusters,
which accrete more rapidly at the periastron than during the rest of the 
binary orbit. They predicted that the repeating ULXB source is likely to be 
periodic if adequate sampling is provided. 
For the ULXBs observed in NGC~5128, \citet{Irwin et al.(2016)} showed that
the fast rise and slow decay of flares seem to be similar to that of type-I 
bursts from Galactic NSs. 
However, the peak luminosities of the ULXBs are 1-2 orders of
magnitude higher than that of type-I bursts, which are typically near the
Eddington limit of an NS. Alternatively, the highly super-Eddington accretion
onto an NS or a stellar-mass BH can satisfy the limitations of peak
luminosities. The flares may be observed at the periastron during the donor
star around the central object in an eccentric orbit 
\citep{Irwin et al.(2016)}. Moreover, \citet{Shen(2019)} showed that the ULXBs
could be explained as the result of accretion onto an IMBH during the
periastron passage of a WD in an eccentric orbit. 

In this Letter, we focus on the repeating ULXB source in NGC 5128 owing to
the detected multiple flares. We first propose an NS-WD binary model with an
eccentric orbit to explain the repeating behavior of the ULXB source in
NGC~5128. The system has a super-Eddington accretion rate. 
At the periastron, the mass transfer occurs when the WD fills its Roche lobe. 
The accreting NSs with super-Eddington
accretion rates may not be rare in the ultraluminous X-ray populations
consisting of NSs \citep[more than 100 times Eddington; e.g.,][]
{Bachetti et al.(2014),Furst et al.(2016),Israel et al.(2017),
Carpano et al.(2018),Quintin et al.(2021)}. 
According to our model, the information of the orbital period may be revealed
by the observed flares.
Thus, we manage to derive the plausible orbital period of the ULXB source
in NGC~5128.
Some methods have been proposed to search for periods when only
a few events were observed \citep[e.g.,][]{Rajwade et al.(2020),Katz(2022)},
which are helpful to derive the plausible orbital period of this ULXB source.

The remainder of this Letter is organized as follows. The NS-WD binary with
a super-Eddington accretion rate is illustrated in Section~\ref{sec:model}.
The plausible periodicity of the ULXB source in NGC 5128 is studied in Section~\ref{sec:period}. The relations between the orbital period, the
eccentricity of the orbit, the WD mass, and the NS mass are shown in Section~\ref{sec:mass}. A possible association between the repeating 
ULXBs and the repeating fast radio bursts (hereafter FRBs) is proposed
in Section~\ref{sec:connection}. Conclusions and discussion are presented 
in Section~\ref{sec:con}.

\section{NS-WD binary model} \label{sec:model}

\begin{figure}[htbp]
\centering
\includegraphics[width=0.8\linewidth]{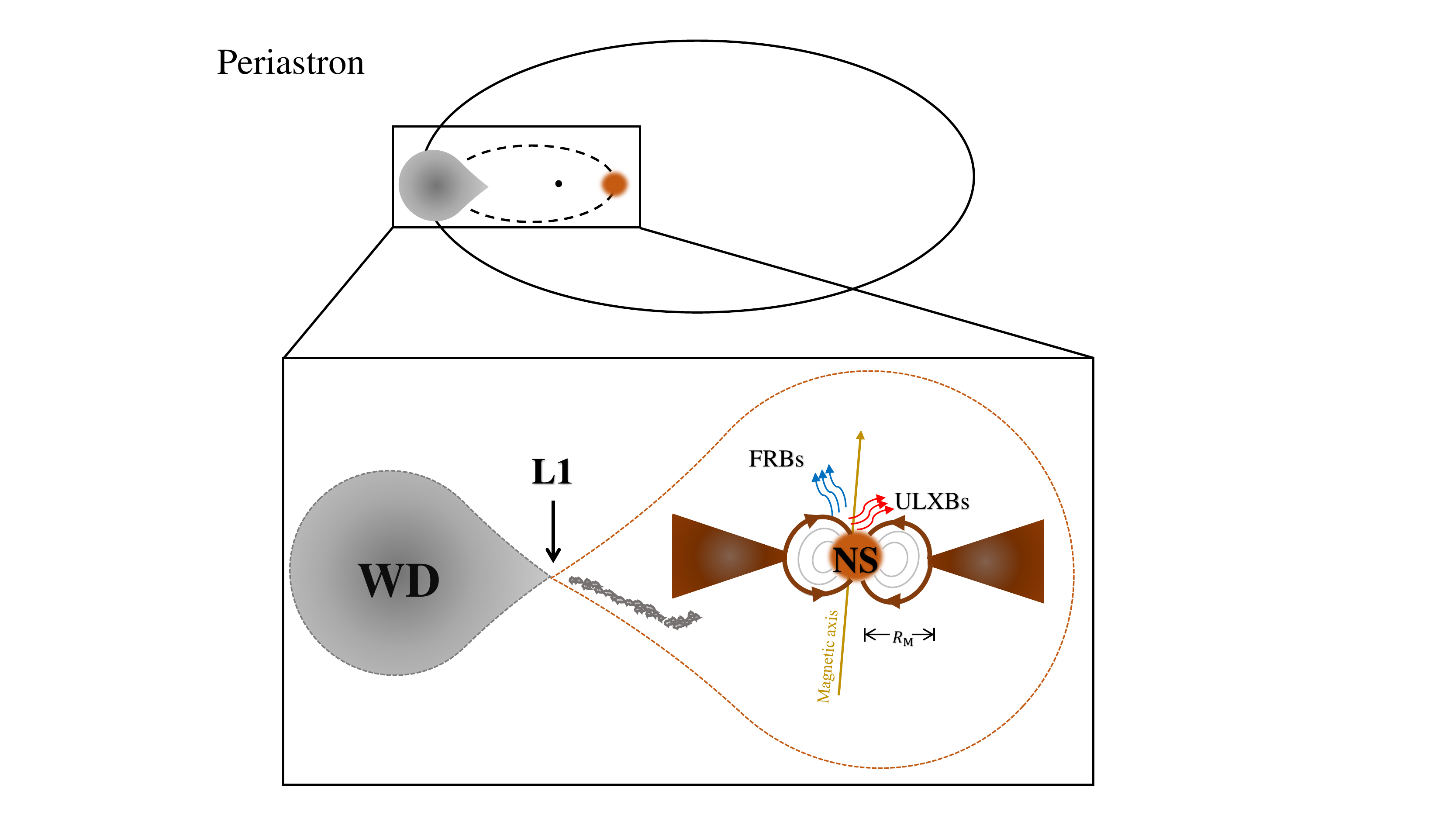}
\caption{The schematic diagram of the NS-WD binary model with an eccentric
orbit, where the WD fills its Roche lobe at the periastron. Around the 
periastron, the mass transfer occurs from the WD to the NS through the inner
Lagrange point $L_{1}$. Due to the viscous processes, the accreted materials
can be fragmented into a number of parts.
The system consists of a magnetic WD and an NS with strong dipole
magnetic fields. The NS accretion disk 
is truncated by the magnetic field near the magnetospheric radius 
$R_{\rm{M}}$. Around the magnetospheric radius $R_{\rm{M}}$, the accreted
material is governed by the magnetic field lines and can leave the disk to 
approach the NS surface. The magnetic reconnection may be triggered. 
The strong FRB radiations can be released by the curvature radiation of the
electrons (more details can be found in Section~\ref{sec:connection}). 
Finally, the accreted materials follow the magnetic field lines down to a 
small area around the polar caps of the NS, forming
an accretion column or funnel. The ultraluminous X-ray 
radiations can be produced from the accretion column.
(The sizes of compact objects are not to scale for the purpose of demonstration.)
\label{fig1} }
\end{figure}

In this section, we propose an NS-WD binary model with a super-Eddington
accretion rate to explain the repeating behavior of the ULXB source in
NGC~5128. The system consists of a magnetic WD and an NS with strong 
dipole magnetic fields. As illustrated in Figure \ref{fig1}, the NS-WD
binary system is in an eccentric orbit.
At the periastron, the mass transfer occurs from the
WD to the NS through the inner Lagrange point $L_{1}$ when the WD fills
its Roche lobe. For other positions on the eccentric orbit, since the Roche
lobe is not filled by the WD, the mass transfer is interrupted. 
Due to the viscous processes, the accreted materials from the WD can be
fragmented into a number of parts.
The accretion flow can be truncated by the magnetic field near the
magnetospheric radius $R_{\rm{M}}$, defined as the location where the magnetic
pressure balances the ram pressure of the accretion flow 
\citep[e.g.,][]{Kaaret et al.(2017)}. Around this location, the magnetized
materials from the WD are governed by the magnetic field lines and leave
the disk to fall onto a small area around the NS magnetic poles \citep[e.g.,][]{1972A&A....21....1P}, yielding an accretion column or funnel. 
The accretion column can be maintained since the radiation pressure is
balanced by the magnetic pressure in the column. The ultraluminous X-ray 
radiations can be produced from the accretion column.
There may be two reasons for the X-ray luminosity of the highly magnetized
NS with an accretion column can exceed the Eddington limit.
One is that the scattering cross-section of the column can be far below the
Thomson value \citep[e.g.,][]{1976MNRAS.175..395B}.
The other reason is that the radiation can escape from the sides
of the column, perpendicular to the incoming flow of the magnetized materials
\citep[e.g.,][]{1988SvAL...14..390L}.
If the NS magnetic field is sufficiently high, 
e.g., $B_{\rm{NS}}\ga 10^{14}\ \rm{G}$, 
the X-ray luminosity can reach the values of the order of 
$10^{40}\ \rm{erg\ s^{-1}}$ \citep{Mushtukov et al.(2015)}, 
which attains the peak luminosities of ULXBs detected in NGC~5128
\citep{Irwin et al.(2016)}.
For the persistent emissions before and after the flares
($L\sim 10^{37} \ \rm{erg \ s^{-1}}$), the collisions of the accreted
materials with the NS disk may play an important role.

On the other hand, \citet{King(2007)} proposed that due to the conservation
of angular momentum, the WD may be kicked away after the Roche-lobe overflow
process for $q<2/3$, where $q$ is defined as the mass ratio of the WD to the 
NS. After that, the gravitational radiation is an essential mechanism to 
enable the WD to refill its Roche lobe. Hence, the NS-WD binary system 
becomes semi-detached again, and the next mass transfers can recur. 
That is, for the binary with $q<2/3$, the Roche-lobe overflow
should be an intermittent process. For a typical NS mass,
$1.4\ M_{\sun}$, the sporadic type of Roche-lobe overflow may be common
for the semi-detached binary system with a WD donor
($M_{\rm{WD}}\la 0.6\ M_{\sun}$).
Therefore, the ULXB is not always observable in each orbital cycle.

\section{Possible orbital periods} \label{sec:period}

\begin{figure}[htbp]
\centering
\includegraphics[width=0.8\linewidth]{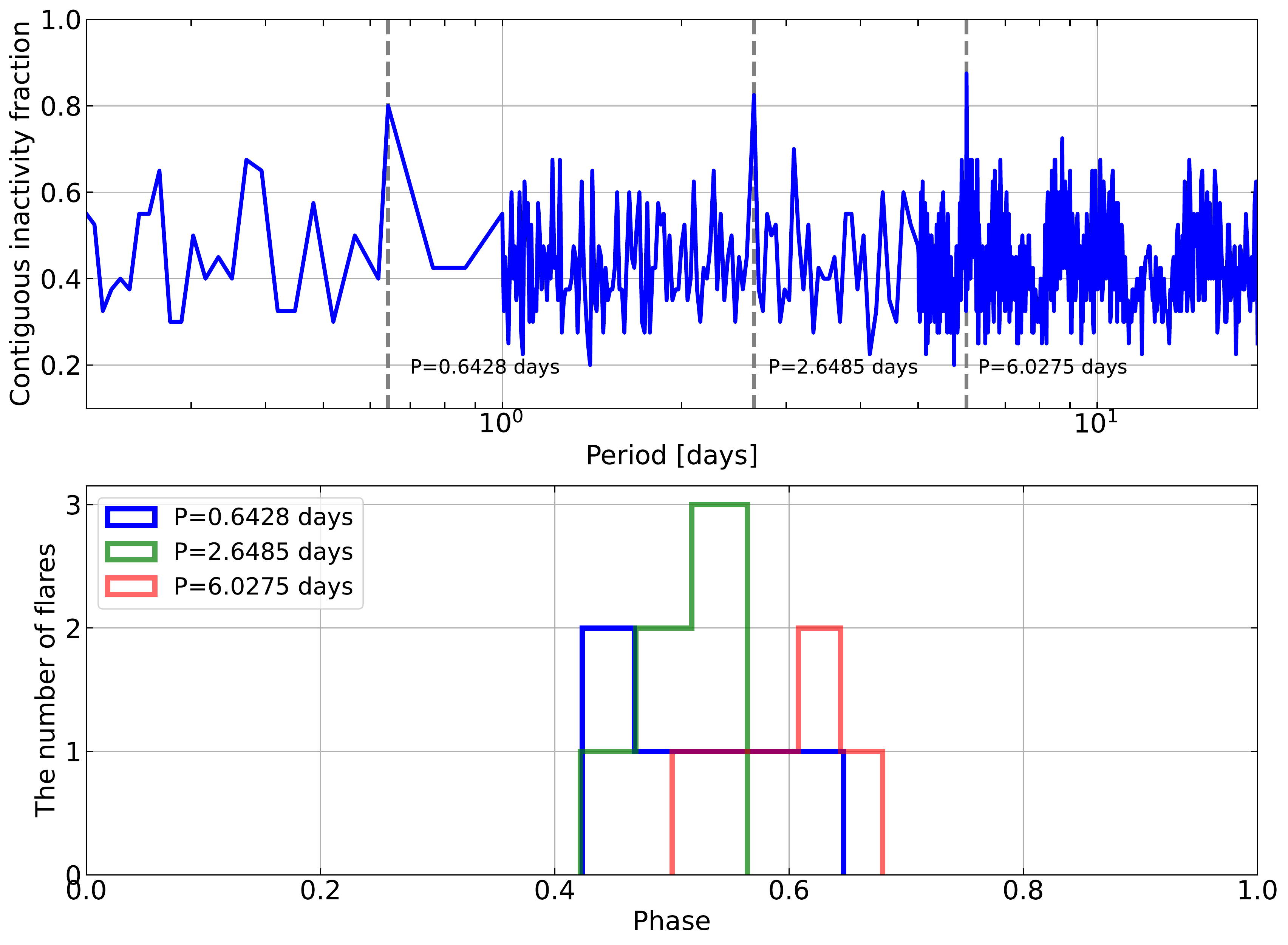}
\caption{Upper panel: Periodogram obtained by folding the arrival time of six
ULXBs at all distinguishable trial periods between $\sim 0.2$ and 
$\sim 18$ days. The folded profiles are evaluated by the length of the
longest contiguous phase without any ULXB activity. Three significant trial
periods can be found, i.e., $P\simeq$ 0.6428, 2.6485, and 6.0275 days.
Lower panel: The activity profiles of the detected ULXBs 
folded with three trial periods. The phase of zero corresponds to the 
beginning time of the first flare minus half a putative period. 
The histograms of the different colors show the distinct periods,
i.e., 0.6428 (blue), 2.6485 (green), and 6.0275 (red) days. It is seen
that an activity window exists for the repeating ULXB in NGC 5128.
\label{fig2} }
\end{figure}

According to the NS-WD binary model described in Section \ref{sec:model},
the orbital period may be revealed by the analyses of the observed
repeating ULXBs. In this section, we manage to derive several possible
orbital periods of the ULXB source in NGC 5128.

We refer to the method of \citet{Rajwade et al.(2020)}, searching for the 
period with the narrowest folded profile. \citet{Rajwade et al.(2020)} folded
the arrival time of FRB 121102 at all distinguishable trial periods between 
2 and 365 days. Then, they measured the length of the longest contiguous
phase without any sourcing activity. Higher values express that the activity
of the FRBs is concentrated within a narrower phase window, which indicates
a periodic activity pattern. According to this method, the trial period for 
FRB 121102 has been found to be $P\simeq159$ days, with an inactive 
contiguous phase of $53\%$ \citep{Rajwade et al.(2020)}. 

For the repeating ULXB source in NGC 5128, five flares were recorded in
\citet{Irwin et al.(2016)}: four were observed with $Chandra$
on March 30, 2007, April 17, 2007, May 30, 2007, and January 4, 2009,
respectively; the fifth flare was observed with $XMM-Newton$
on February 9, 2014.
Another possible flare from the same location of NGC 5128 was observed
by $Swift$ on July 31, 2013 (the information is originally from the talk
of Jimmy A. Irwin in AAS HEAD Meeting in August 2017, and also from some private
communications with Jimmy A. Irwin).
$Swift$ observed seven photons in 300~s (the total observation time of
$\sim 3400$ s). 
The count rate of the flare 
($\sim 0.023 \ \rm{count \ s^{-1}}$; time bin of 300 s) is far higher than
the count rates before and after the flare
($\sim 0.001 \ \rm{count \ s^{-1}}$). 
Thus, it may be regarded as another flare from the same ULXB source in NGC 5128.

Based on the beginning time of six flares, the minimum interval between two 
adjacent detected flares is $\sim18$ days, which should be an upper limit
of the orbital period of the ULXB source in NGC~5128.
Hence, based on the method of \citet{Rajwade et al.(2020)}, we folded the
beginning time of six flares at all distinguishable trial periods between 
0.2 days and $\sim 18$ days. The fraction of ULXB inactivity as a
function of the period is shown in the upper panel of Figure~\ref{fig2}. Three 
significant trial periods can be found, i.e., 0.6428~days, 2.6485~days, and 
6.0275~days, which display the inactive contiguous phases of $80\%$, $83\%$, 
and $88\%$, respectively.
By folding the beginning time of six flares with three trial periods, the
activity profiles of the ULXBs are shown in the lower panel of 
Figure~\ref{fig2}. The histograms of different colors exhibit the distinct
plausible activity windows of the repeating ULXB source in NGC~5128.

Summing up all the available data (up to May 17, 2015), 
\citet{Irwin et al.(2016)} evaluated the recurrence rate (defined as the total
observation time divided by the number of flares) of one flare every 
$\sim 1.8$ days (a total observation time of $7.9\times 10^{5}$ s divided 
by five flares) and the duty cycle of $\sim 2.5\%$ for the repeating ULXB 
source in NGC 5128. 
For now, six flares were observed in a total observation time of 
$\sim 1.2\times 10^{6}$~s, yielding the recurrence rate of one flare every 
$\sim 2.3$ days. Based on our model, since the Roche-lobe overflow should 
be an intermittent process, the ULXB is not always observable in each
orbital cycle. Thus, the orbital period of the ULXB source in NGC~5128
is likely $P_{\rm orb} \la 2.3\ \rm{days}$.
Thus, among the three possible orbital periods, 0.6428 and 2.6485 days
are more preferred.
We would point out that, only one period can be correct and
two wrong periods show up, which indicates that only six observed flares
are not sufficient to well constrain the orbital period.

\section{Masses of the compact objects} \label{sec:mass}

\begin{figure}[htbp]
\centering
\includegraphics[width=0.5\linewidth]{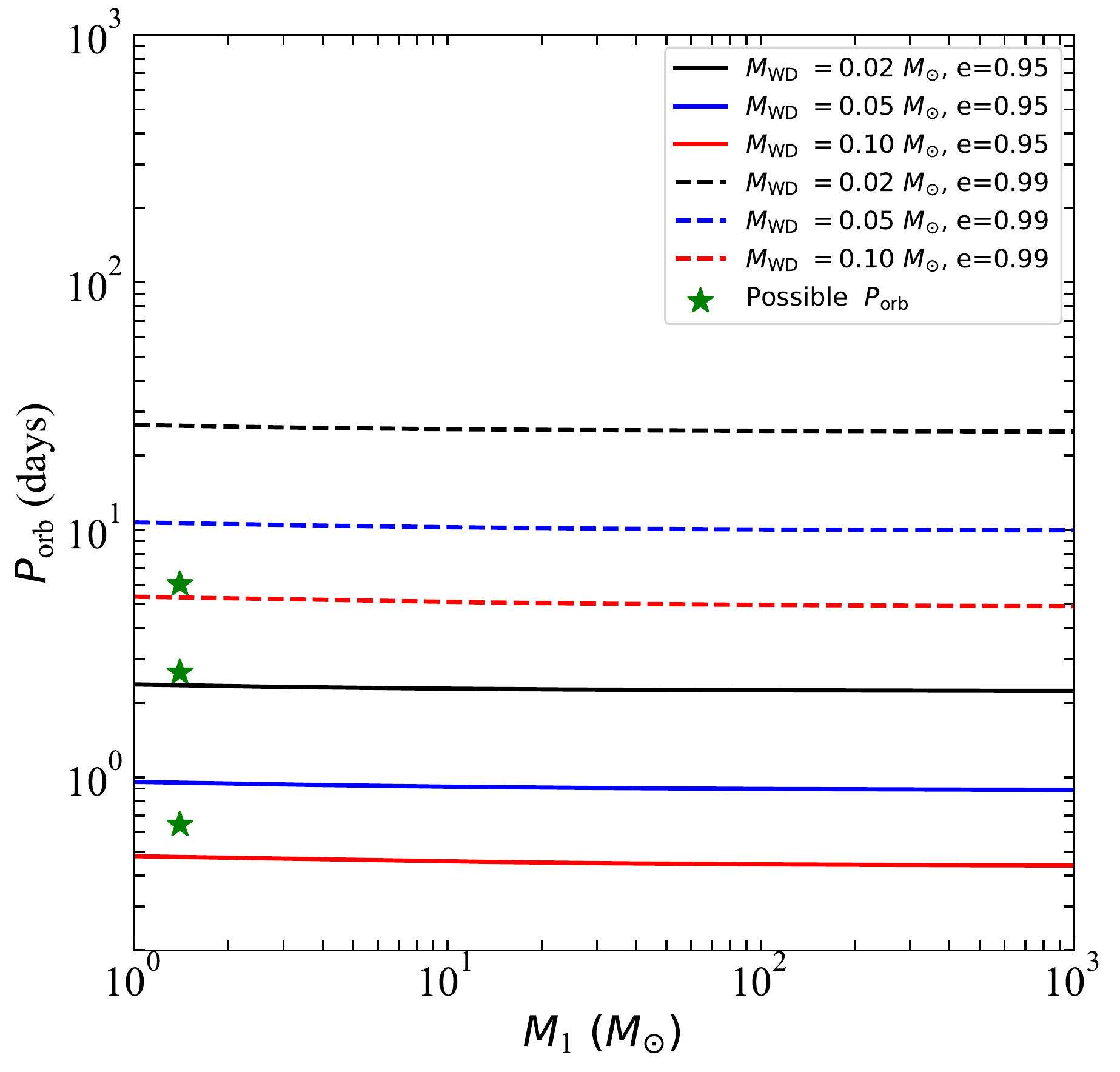}
\caption{The relationship between the orbital period $P_{\rm{orb}}$ and 
the mass of the central object $M_{1}$ for the given eccentricities $e$ and 
the WD masses $M_{\rm{WD}}$. The lines of different colors represent distinct
WD masses: $0.02\ M_{\sun}$ (black), $0.05\ M_{\sun}$ (blue), and 
$0.1\ M_{\sun}$ (red). The solid and dashed lines represent $e=0.95$ and 0.99, respectively. The green stars represent three possible orbital periods of
the repeating ULXB source in NGC 5128, assuming that $M_{1}=1.4\ M_{\sun}$.
\label{fig3} }
\end{figure}

Three possible orbital periods of the repeating ULXB source in NGC 5128 have
been found in Section~\ref{sec:period}. We consider that the orbital periods
are related to the orbital period of the NS-WD binary. In addition to the 
explanation of the super-Eddington accretion of an NS, the ULXBs can also
be considered to be produced by the sub-Eddington accretion of an IMBH.
\citet{Shen(2019)} proposed that ULXBs can be explained by the 
accretion onto an IMBH during the periastron passage of a WD in an eccentric
orbit. Moreover, they predicted that the interval between two recurrent flares 
may be the eccentric orbital period of the WD. A large eccentricity $e$ of 
the orbit ought to be considered to satisfy the restriction of the long 
orbital period \citep[e.g., $P\sim 1~\rm{day}$, $e \sim 0.97$; 
see Equation (8) in][]{Shen(2019)}. 
For the binary systems with different types (or masses) of central objects, 
i.e., NS or IMBH, in the following section, we try to confirm whether the
periodicity of the ULXB source in NGC~5128 can be affected.

The dynamic equation of the binary system can be expressed as 

\begin{equation}
\frac{G \left(M_{1}+M_{\rm{WD}} \right)}{a^{3}}=\frac{4 \pi^{2}}
{P_{\rm{orb}}^{2}}
\label{e2}\ ,
\end{equation}
where $G$ is the gravitational constant, $M_{1}$ and $M_{\rm{WD}}$ are
the masses of the central object (i.e., NS or IMBH) and the WD, respectively, 
$a$ is the binary separation, and $P_{\rm{orb}}$ is the orbital period. 
The Roche-lobe radius $R_{\rm{L}}$ for the WD at the periastron can 
take the form \citep{Eggleton(1983)}:

\begin{equation}
\frac{R_{\rm{L}}}{a \left(1-e \right)}=\frac{0.49 q^{2/3}}{0.6q^{2/3}+\ln \left(1+q^{1/3}\right)} \ ,
\label{e3}
\end{equation}
where $q$ is the mass ratio defined as $q=M_{\rm{WD}}/M_{1}$. The WD radius $R_{\rm{WD}}$ can be expressed as \citep{Tout et al.(1997)}

\begin{equation}
R_{\rm{WD}}=0.0115 R_{\sun} \sqrt{\left(M_{\rm{Ch}}/M_{\rm{WD}}\right)^{2/3}-
\left(M_{\rm{WD}}/M_{\rm{Ch}}\right)^{2/3}} \ ,
\label{e4}
\end{equation}
where $M_{\rm{Ch}}$ is the Chandrasekhar mass limit 
$M_{\rm{Ch}}=1.44 \ M_{\sun}$. According to the assumption that the WD fills 
its Roche lobe at the periastron, i.e., $R_{\rm{WD}}=R_{\rm{L}}$,
once $M_{1}$, $M_{\rm{WD}}$, and $e$ are given, 
the orbital period can be derived by Equations (\ref{e2})-(\ref{e4}). The
relationships between the orbital period $P_{\rm{orb}}$ and the mass of 
central object $M_{1}$ for the distinct given eccentricities $e$ and the WD
masses $M_{\rm{WD}}$ are shown in Figure~\ref{fig3}.
The lines of different colors represent different masses of the WDs, i.e., 
$0.02\ M_{\sun}$ (black), $0.05\ M_{\sun}$ (blue), and $0.1\ M_{\sun}$ (red). 
Moreover, the solid and dashed lines express $e=0.95$ and $0.99$, 
respectively. It is seen from Figure~\ref{fig3} that once the mass of
the WD $M_{\rm{WD}}$ and the eccentricity of the orbit are determined, the 
orbital period $P_{\rm{orb}}$ of the system nearly remains constant for
varying $M_{1}$. In other words, for such an accretion model,
a large eccentricity is required to interpret the repeating ULXB source
in NGC~5128 no matter the accretor is an NS, a stellar-mass BH, or an IMBH.
For the possible orbital periods of this source 
(the green stars shown in Figure~\ref{fig3}), $e>0.95$ is required
for $M_{\rm{WD}}=0.1 \ M_{\sun}$ and $M_{\rm{1}}=1.4 \ M_{\sun}$.

In addition, a simple analytic relation between $P_{\rm{orb}}$, $M_{\rm{WD}}$
and $e$ can refer to Equation (7) of \citet{Gu et al.(2020)}.
Such a simple relation is independent of $M_{1}$. Hence, it is reasonable
that the mass of the central object $M_{1}$ cannot validly influence the 
orbital periods of the eccentric binary systems. That is, the orbital period 
of the repeating ULXB source in NGC 5128 is mainly related to the mass of the
WD and the eccentricity of the orbit. 

\section{Possible association between ULXBs and FRBs} \label{sec:connection}

In our previous studies, the NS-WD binary model with super-Eddington 
accretion rates has been proposed to explain the repeating FRBs. 
In that model, the repeating FRBs can be explained as the result of the
magnetic reconnection when the accreted magnetized materials approach the NS
surface \citep{Gu et al.(2016),Gu et al.(2020),Lin et al.(2022)}, or the radio
radiations of the narrowly collimated jet \citep{Chen et al.(2021)}. 
FRBs are millisecond-duration radio bursts with the isotropic equivalent
luminosities of $\sim 10^{38-42} \ \rm{erg\ s^{-1}}$ 
\citep[for reviews, see][]{2019ARA&A..57..417C,Petroff et al.(2019)}. 
Recently, FRB~20200120E was discovered to be located in a globular cluster
associated with M81 \citep{Kirsten et al.(2022)}.
This source is the first detected
repeating FRB located in an extragalactic globular cluster, whereas other
located repeating FRBs are related to the nearby star-forming regions of 
the host galaxies \citep[e.g.,][]{Bassa et al.(2017),Tendulkar et al.(2021)}.
No X-ray source at the location of FRB 20200120E has been found in archival
$Chandra$ observations, which results in a 0.5-10 keV luminosity upper 
limit of $2\times 10^{37} \ \rm{erg \ s^{-1}}$ at the distance of 3.6 Mpc
\citep{Kirsten et al.(2022)}.
\citet{Kirsten et al.(2022)} proposed that the
ultraluminous X-ray sources associated with FRB 20200120E
can be ruled out by the X-ray limit unless the X-ray luminosities of these
sources vary in time by more than two orders of magnitude.
Based on the burst arrival times of FRB 20200120E, \citet{Nimmo et al.(2022b)} 
found a possible 12.5-day activity period, which may require more
detected bursts to test the significance of this period.
On the other hand, in our opinion, the repeating ULXB source has
the possibility to be associated with the repeating FRB source
owing to its extremely violent burst ability.
The repeating ULXB source and the repeating FRB source
show the most violent bursts in the X-ray and the radio bands, respectively.
In addition, our NS-WD binary model with super-Eddington accretion rates
can work well for both of these two systems. 
Moreover, \citet{Katz(2021)} proposed that an NS and a close
binary companion (likely a WD) can be responsible for the observed
long-lived repeating FRB sources in globular clusters.
Thus, we propose that a possible association may exist
between the repeating ULXB source and the
repeating FRB source. Such an association is worth further investigation
by follow-up observations on nearby extragalactic globular clusters.

The burst timescales of $\sim 60 \ \rm{ns}$ to $5\ \rm{\mu s}$ observed
in FRB~20200120E indicate a light-travel size of $\sim$ 20-1500 m 
\citep{Majid et al.(2021)}, which supports a magnetospheric origin of the 
FRB radiations \citep[e.g.,][]{Zhang(2020)}. \citet{Nimmo et al.(2022a)} 
proposed that the magnetic reconnection events in the close vicinity of a 
highly magnetized NS can produce the observed timescales and luminosities
from FRB 20200120E. Thus, it is reasonable that the system consists of a
magnetic WD and an NS with strong dipole magnetic fields. 
The schematic diagram of the general physical process is shown in 
Figure~\ref{fig1}.
Around the magnetospheric radius $R_{\rm{M}}$, the magnetized accreted 
materials are guided by the magnetic field lines and leave the disk to follow
the magnetic field lines down to the polar caps of the NS. 
During the falling process, the magnetic reconnection may
be triggered. The strong electromagnetic radiations can be released by the
curvature radiation of the electrons, which move along the NS magnetic field
lines with ultra-relativistic speed 
\citep[e.g.,][]{2011ApJ...726...90Z,Gu et al.(2016)}. 

According to our model, the duration of an FRB may be regarded as the
timescale of  a magnetic reconnection. The duration $t_{\rm{d}}$ can be 
evaluated by the ratio of the NS radius $R_{\rm{NS}}$ to the
Alfv$\rm{\acute{e}}$n speed $v_{\rm{A}}$
\citep[=$B_{\rm{NS}}/\sqrt{4\pi \bar{\rho}}$ ;][]{Gu et al.(2016)} 

\begin{equation}
t_{\rm{d}}=\frac{B_{\rm{NS}}}{v_{\rm{A}}}=1.1 \left(\frac{R_{\rm{NS}}}{10^{6}\ \rm{cm}}\right) \left(\frac{B_{\rm{NS}}}{10^{11}\ \rm{G}}\right)^{-1} \left(\frac{\bar{\rho}}{10^{3}\ \rm{g\ cm^{-3}}}\right)^{1/2}\ \rm{ms}\ ,
\end{equation}
where $B_{\rm{NS}}$ is the magnetic flux density of the NS, and
$\bar{\rho}$ is the averaged mass density of accreted materials. For the 
typical values $R_{\rm{NS}}=10^{6}\ \rm{cm}$
and $\bar{\rho} =10^{3}\ \rm{g\ cm^{-3}}$, 
a large magnetic flux density ($B_{\rm{NS}}\ga 10^{14}\ \rm{G}$)
is required to account for the microsecond-duration bursts of FRB 20200120E
\footnote{For the mass density in the atmosphere of a WD,
continuous variation exists from large interior values to essentially zero.
For simplicity, following \citet{Gu et al.(2016)}, we use an averaged mass
density $\bar{\rho}=10^{3}\ \rm{g\ cm^{-3}}$ for the accreted materials.}.

\section{Conclusions and discussion} \label{sec:con}

In this Letter, we have proposed an NS-WD binary model with an eccentric 
orbit to explain the repeating behavior of the ULXB source in NGC~5128. 
The system has a super-Eddington accretion rate.
Moreover, for the most violent short-duration repeating bursts in the X-ray
and radio bands, i.e., the repeating ULXBs and the repeating FRBs such as 
FRB 20200120E, which were detected in extragalactic globular clusters, we
propose that the physical processes of these two types of bursts may have a 
possible association. Our NS-WD binary model with super-Eddington
accretion rates can account for this association. 
According to our model, the orbital period may be revealed by the
observed repeating bursts of the ULXB source in NGC 5128.
We have derived three possible periods by analyses of the detected bursts,
i.e., 0.6428, 2.6485, and 6.0275 days, where the former two periods
are more preferred. Whether the central object is an NS or an IMBH, 
a large eccentricity of the orbit is required to interpret the orbital periods,
e.g., $e>0.95$ for $M_{\rm{WD}}=0.1 \ M_{\sun}$.

\citet{Li et al.(2021)} showed that 1652 independent bursts of FRB 121102
have been detected by FAST in 59.5 hours spanning 47 days.
The burst rate of FRB~121102 peaked at 122 $\rm{h^{-1}}$, which dropped
dramatically afterward. The variations of the burst rates indicate the changes
in the activities of FRB~121102. As the first discovered repeater,
the activity timescale of FRB~121102 may be evaluated as $\ga 10 \ \rm{years}$.
According to our model, the activity timescale of the ULXB source in NGC 5128
should be similar to that of the repeating FRBs, i.e., $\ga 10 \ \rm{years}$.
For the ULXB source in NGC 5128, six flares were observed from 2007 to 2014. 
After that, no significant flares were discovered from the same location,
implying that the $\ga 10$-year activity timescale for the ULXB source
in NGC 5128 is reasonable.

According to our model, a large eccentricity $e$ of the orbit is required to
interpret the plausible orbital period of the ULXB source in NGC~5128. 
There are two possibilities for forming an NS-WD binary with a large
eccentricity. The first is caused by the natal kick of the supernova
explosion in a binary channel, which may correspond to a young NS.
The other possibility is that a WD can be tidally captured by an old 
NS \citep[e.g.,][]{Clark(1975)}. The globular clusters host old stellar 
populations and have extreme stellar densities.
Thus, in the globular clusters, the second possibility
should play an essential role in the formation of an NS-WD binary with a
large eccentricity.

The distance of NGC 5128 is 3.8 Mpc \citep{Harris et al.(2010)}, and the 
distance of M81 is 3.6 Mpc \citep{Freedman et al.(1994)}. 
The compact binaries are efficiently formed inside the globular clusters of 
these nearby extragalactic galaxies. 
Follow-up observations on nearby extragalactic globular clusters may help
to examine the potential association between the repeating ULXBs and the 
repeating FRBs.

\acknowledgments

We thank Jimmy A. Irwin and Di Li for beneficial discussions,
and thank the anonymous referee for helpful comments that improved the paper.
This work was supported by the National Key R\&D Program of China under
grant 2021YFA1600401, and the National Natural Science Foundation of China
under grants 11925301, 11973002, 12033006, U1831205, and U2038103.

\end{document}